# Heterogeneously Integrated Memristive Laser on Silicon with Non-Volatile Wavelength Tuning

Bassem Tossoun*, Di Liang, Xia Sheng, John Paul Strachan and Raymond G. Beausoleil
Hewlett Packard Labs, Hewlett Packard Enterprise, 820 N McCarthy Blvd, Milpitas, CA 95305, USA
*e-mail: bassem.tossoun@hpe.com

**The von-Neumann bottleneck has constrained computing systems from efficiently operating on the increasingly large demand in data from networks and devices. Silicon (Si) photonics offers a powerful solution for this issue by providing a platform for high-bandwidth, energy-efficient interconnects. Furthermore, memristors have emerged as a fundamental building block for non-volatile data storage and novel computing architectures with powerful in-memory processing capabilities. In this paper, we integrate an $Al_2O_3$ memristor into a heterogeneous Si quantum dot microring laser to demonstrate the first laser with non-volatile optical memory. The memristor alters the effective optical modal index of the microring laser cavity by the plasma dispersion effect in the high resistance state (HRS) or Joule heating in the low resistance state (LRS), subsequently controlling the output wavelength of the laser in a non-volatile manner. This device enables a novel pathway for future optoelectronic neuromorphic computers and optical memory chips.**

Over the past several decades, computer processors' performance progressed in parallel to Moore's Law as transistor density increased and multi-core processors were developed[1]. However, within the last decade, processing performance have plateaued as leakage power increase with transistor density and processing latency is limited at a maximum number of parallel processing cores, as Amdahl's law explains[2]. Moreover, the von-Neumann bottleneck caps the amount of data that can be transferred from the processor to the memory, and the gap between memory and processor performance exacerbates this deficiency even further[3]. These roadblocks delay the technological pace towards the development of an Exascale supercomputer, which will be invaluable to keep up with our society's rising data and computing demands from Internet of Things (IoT), autonomous driving, machine learning and artificial intelligence.

However, several technological breakthroughs over the last two decades have opened up novel opportunities to battle these challenges. For instance, Si photonics offers a promising technology platform for low-cost, dense-integration, high-bandwidth, energy-efficient optical interconnects within data centers and high-performance computers (HPC)[4]. The technology has been largely developed over the last decade to integrate all types of active and passive photonic devices into photonic integrated circuits made of Si, much like the evolution of the microelectronic integrated circuit. In particular, the heterogeneous III-V-on-Si platform enables a seamless integration between gain sources (lasers and amplifiers) and Si photonic circuits and shows the potential for high integration in a volume manufacturing process[5]. On the other hand, the memristor, a nanoscale electronic device, was experimentally demonstrated in 2008[6]. By applying a high electric field across the oxide layers of the memristor, oxygen atoms can ionize and leave behind vacancies which act as a conductive path for current flow in the oxide layer, effectively changing the resistance of the oxide material. These devices became an attractive candidate for fast, low-power, non-volatile memory to replace mainstream Flash memory. They also became useful as the fundamental elements of memristive crossbar arrays and artificial neural synapses[6].

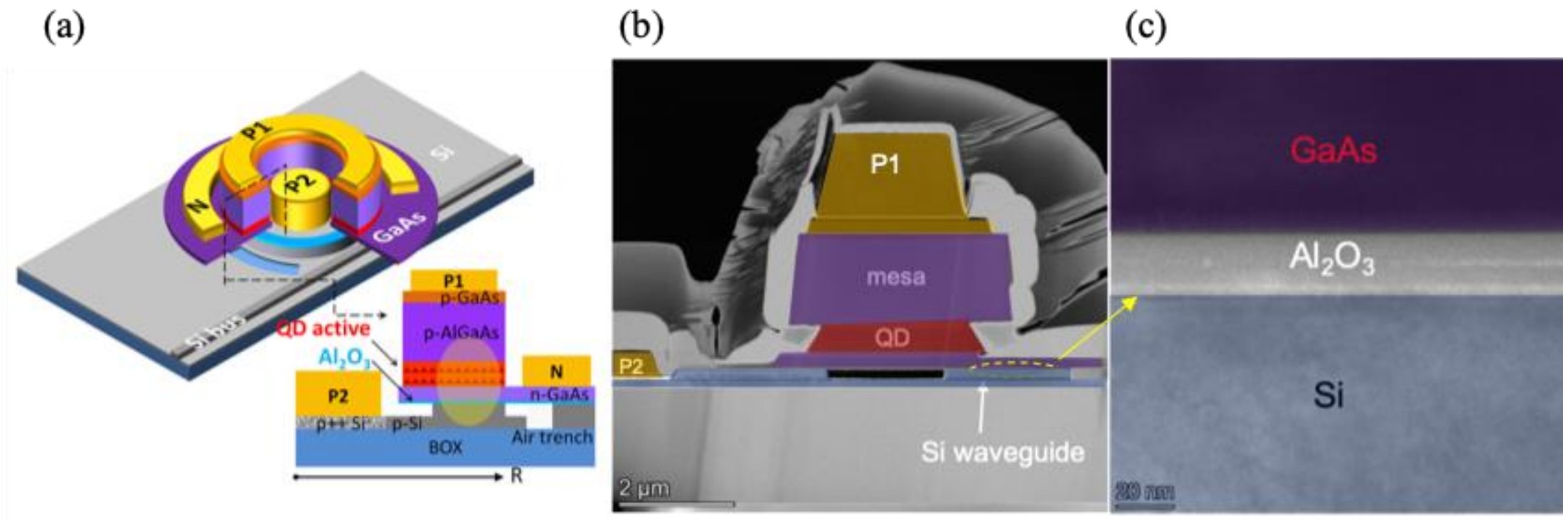

**Fig. 1 | Schematic diagrams of a heterogeneous microring laser integrated with a memristor. a.** 3D view and (half) cross-sectional view with simulated fundamental TE mode and memristor region (dash line box). **b.** Transmission electron microscopy (TEM) image of memristor laser after FIB. **c.** TEM image of a bonded GaAs-to-Si structure after atomic layer deposition (ALD) and direct wafer-bonding process[9].

In this work, we integrate an $Al_2O_3$ memristor within a heterogeneous III-V-on-Si microring laser to realize the memristor laser, the first-ever laser diode with non-volatile optical memory. A three-dimensional schematic of the device as well as a cross-section schematic can be seen in Fig. 1(a). The device comprises an Indium Arsenide (InAs)/Gallium Arsenide (GaAs)-based quantum dot (QD)-based laser sitting on top of a 50-µm diameter Si microring resonator[8]. About 10 nm of $Al_2O_3$ was grown on both a III-V sample and a Si wafer using atomic layer deposition (ALD) and then they were bonded together to form a heterogeneous structure[9]. The three-terminal heterogeneous laser geometry is shown with a common ground terminal N, a P1 terminal which biases the p-contact of the laser, and a P2 terminal which biases the highly-doped p++ Si region. Electrons and holes are injected through the terminals N and P1, respectively, and recombine in the QD active region. A Si bus waveguide is placed adjacent to the microring to evanescently couple some light out of the laser cavity. The newly added third terminal (P2) resides on the highly-doped P++ Si layer in the center of the microring laser.

As highlighted in the cross-sectional transmission electronic microscopy (TEM) images in Fig. 1(b) and (c), respectively, the Si layer and n-GaAs layer sandwich the ~20 nm-thick layer of $Al_2O_3$ to form a semiconductor-insulator-semiconductor (SIS) structure memristor. Due to fabrication imperfections, there was significant undercut of the QD laser active region, which degraded the laser efficiency, but did not impact the SIS memristor. An air trench is etched through the top Si layer for electrical and optical isolation. It also confines the SIS structure to the fundamental TE lasing mode area to efficiently apply the memristive effect on the laser operation and effectively eliminate the high-order transverse modes in the heterogeneous resonator.

After applying a high positive bias voltage across the SIS memristor, an electroforming process takes place inside the oxide material, causing oxygen atoms to migrate and leave behind vacancies which are negatively ionized and creating a conductive path for current to flow, effectively changing the conductivity of the oxide material from a high-resistance state (HRS) to a low-resistance state (LRS)[10]. In the LRS, the device is effectively converted from behaving as a capacitor to behaving as a resistor. Afterwards, it can be reset to the HRS by applying a high electric field across the oxide layer in the opposite direction, eliminating the conductive path that was previously formed such that it now behaves predominantly as a capacitor. When a positive bias is applied again, the conductive path reconnects and the device is reset to the LRS, effectively converting the device from a capacitor to a resistor. Prior studies suggest that both electric field and Joule heating can play a significant role in the physical switching mechanism of $Al_2O_3$-based memristors[11].

This was confirmed through simulations in which a 15 nm × 15 nm × 20 nm filament of aluminum was inserted in the $Al_2O_3$ layer to act as a conductive path for current and heat to flow. The dimensions the aluminum filament was chosen to replicate conductive filaments formed in similar devices[12]. The simulations show that the conductive path created localized Joule heating in the memristor, which increases the effective modal index of the laser cavity. Fig. 2(a) is an optical simulation showing the resonator's fundamental TE lasing mode overlapping the SIS region so that the laser threshold current, output power, and lasing wavelength will change as function of input voltage bias. Fig. 2(b) shows the simulated change in refractive index of the waveguide and phase shift versus applied voltage bias across of the memristor in the HRS and LRS state. In Fig. 2(c), the temperature profile of the device is simulated while the memristor is in the LRS and biased at 5 V.

When a low voltage is applied across the memristor in the HRS, the free carrier absorption (FCA) loss and the effective index of refraction in the waveguide decreases due to the plasma dispersion effect in the Si and GaAs layers of the waveguide. Without the bias applied on terminal P2, the lightly doped p-Si ($1 \times 10^{17}$ $cm^{-3}$) is largely depleted because of the highly doped n-GaAs ($3 \times 10^{18}$ $cm^{-3}$) on the other side of the waveguide, introducing negligible background FCA in p-Si. Once a forward bias is applied to P2, free holes and electrons quickly accumulate in the Si and n-GaAs, respectively, their concentrations increases exponentially when approaching the oxide/Si interface. The output lasing wavelength and output power are both reduced as a result of a corresponding reduction in effective modal refractive index and increase in cavity internal loss[13].

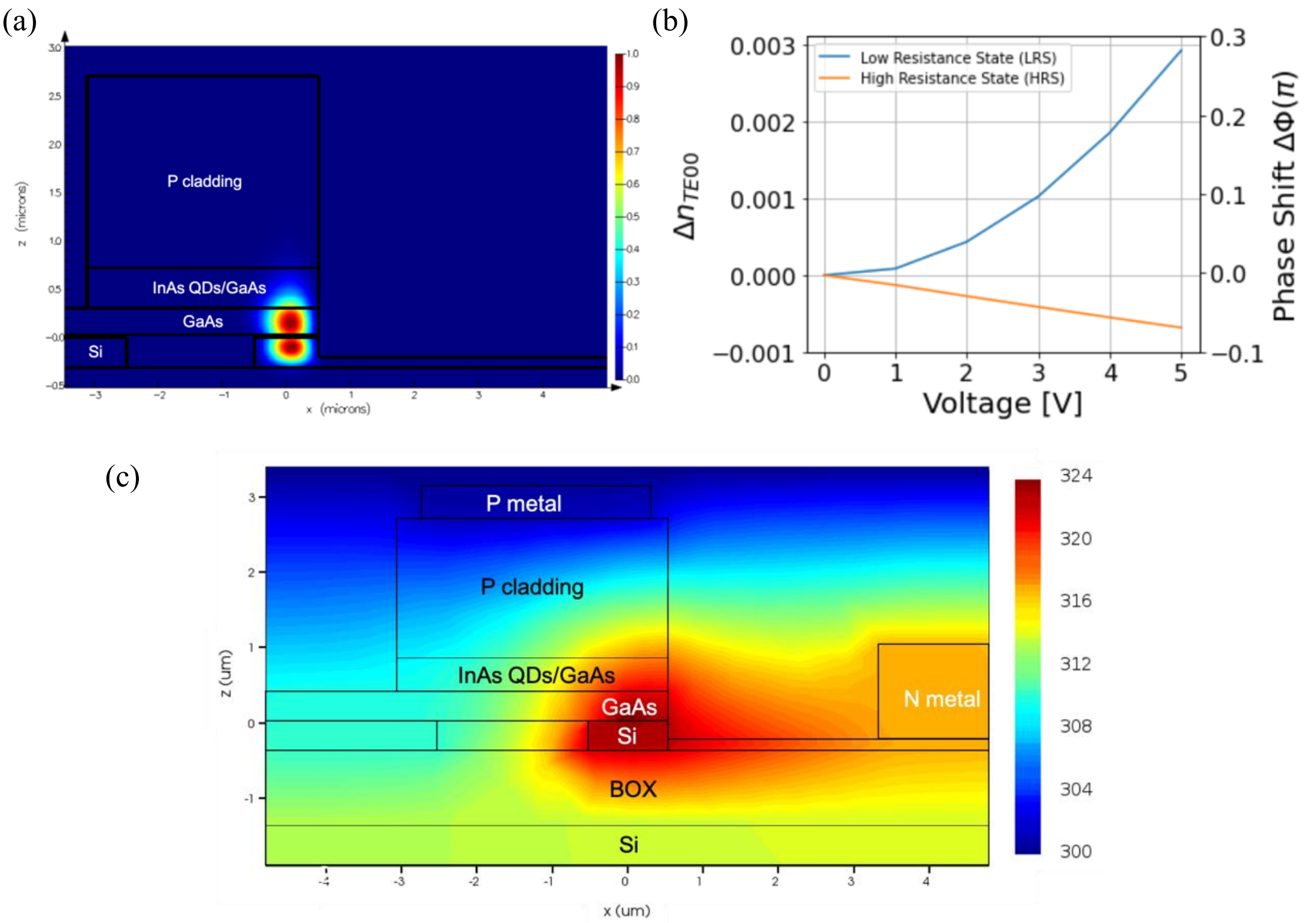

**Fig. 2 | Device simulations. a.** Simulated fundamental TE mode inside microring waveguide, as well as, corresponding calculations of and **b.** refractive index and phase shift versus applied voltage bias of the memristor laser with a 50 μm-diameter in the HRS and the LRS. **c.** Simulated temperature profile of device while the memristor laser is in the LRS and biased at 5 V and temperature color map in Kelvins on the right.

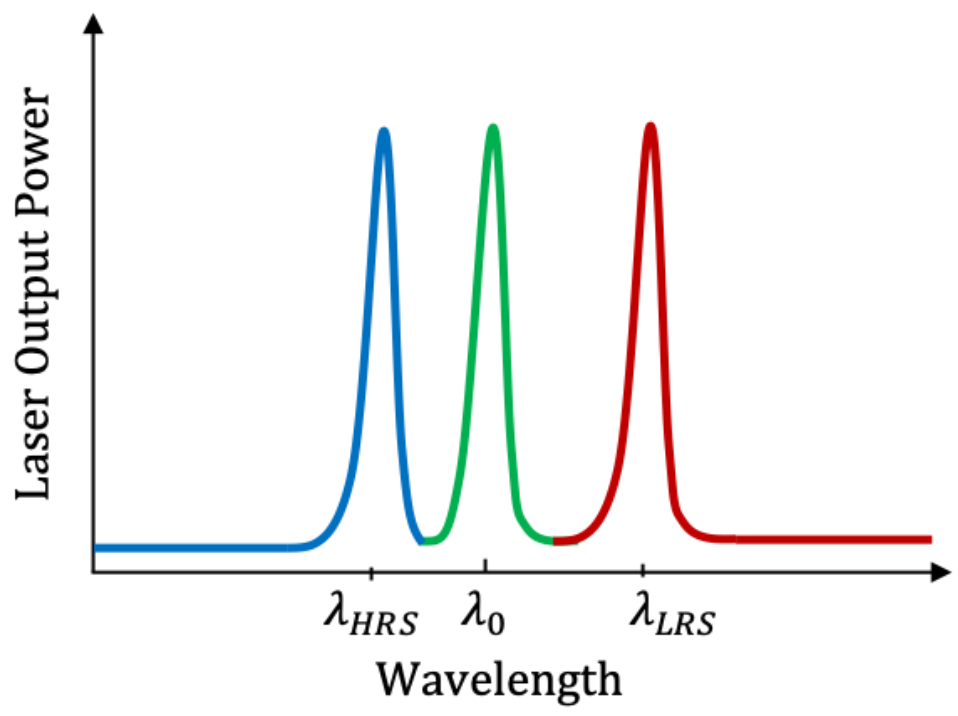

**Fig. 3 |** Memristor laser output power as a function of wavelength in its original state, the LRS and the HRS.

When the memristor is the LRS, localized Joule heating dominates and the plasma dispersion effect diminishes, which increases effective refractive index and causes a red-shift in wavelength. The change in effective index is also about 3-4× larger in the LRS of the memristor than the HRS of the memristor because Joule heating is a stronger effect than the plasma dispersion effect in this case. By using a thinner oxide with a higher dielectric constant, such as $HfO_2$ or $TiO_2$, one could achieve a stronger plasma dispersion effect and tuning range in the HRS of the device[14].

As shown in Fig. 3, we denote the original lasing wavelength before the electroforming process is done in the memristor, as $\lambda_0$. Switching the memristor between the LRS and the HRS subsequently switches the output wavelength of the memristor laser between two characteristic wavelengths: $\lambda_{HRS}$, the lasing wavelength shorter than $\lambda_0$ with a positive bias applied to the memristor in the HRS, and $\lambda_{LRS}$, the lasing wavelength longer than $\lambda_0$ with a positive bias applied to the LRS. Thus, a binary bit of data representing a 0 or a 1 can be electrically stored by setting the memristor to the HRS or LRS, and then the stored data can be optically read by measuring the laser wavelength with respect to the original lasing wavelength. Since this process is non-volatile, data can be stored with zero power being delivered to the laser. For instance, after setting the memristor to the LRS, the laser operates at the output wavelength, $\lambda_{LRS}$. Then, after removing the input bias voltage and power to the laser and then reapplying it, the output wavelength of the laser remains unchanged at $\lambda_{LRS}$. After resetting the memristor to the HRS and applying the same positive bias to the memristor, the laser output wavelength then switches back to $\lambda_{HRS}$ until the memristor is switched back to the LRS. Furthermore, multiple resistance levels have been observed for similar devices by varying the input voltage bias[15]. Multiple low resistance states then produce multiple output lasing wavelengths, $\lambda_{LRS_{1-n}}$, enabling multiple bits within the device for higher data storage.

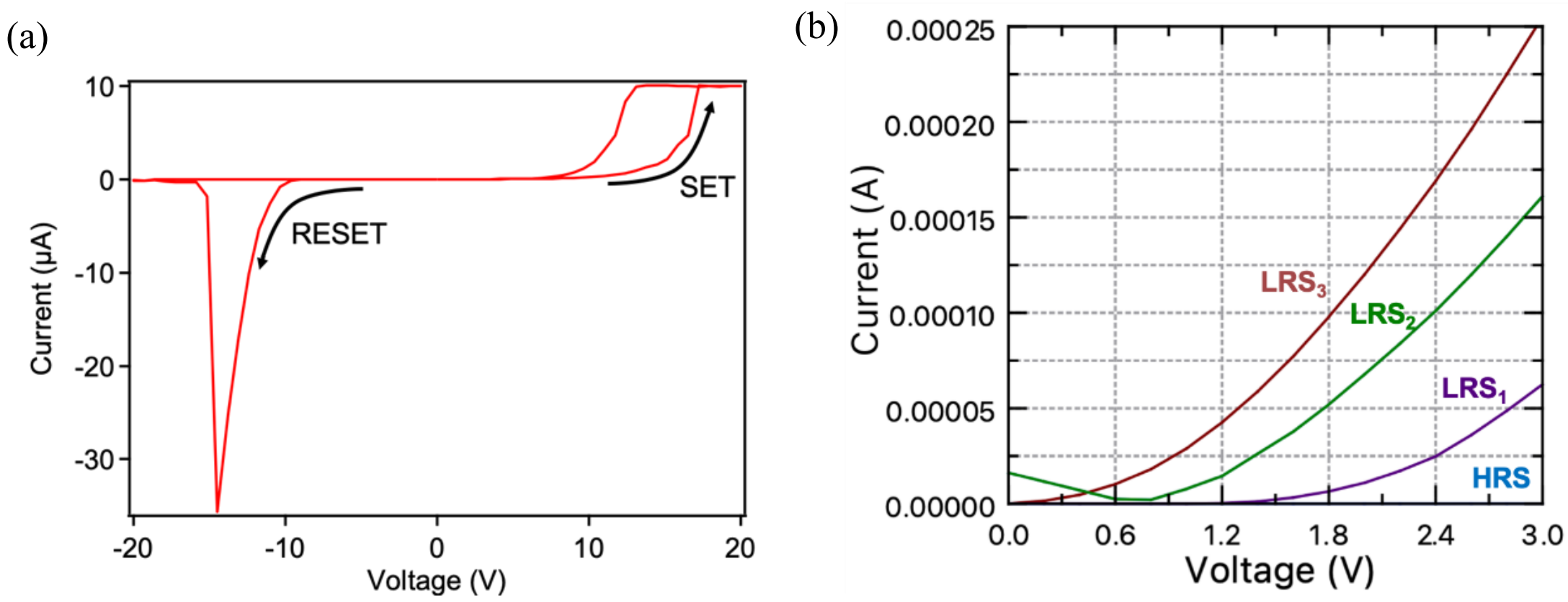

**Fig. 4 | a.** I-V curve of memristor being set to LRS around 16 V and then reset to the HRS around -14 V. Hysteresis loop observed, confirming memristor-like behavior. **b.** Measured switching speed of the memristor being switched from the HRS to the LRS.

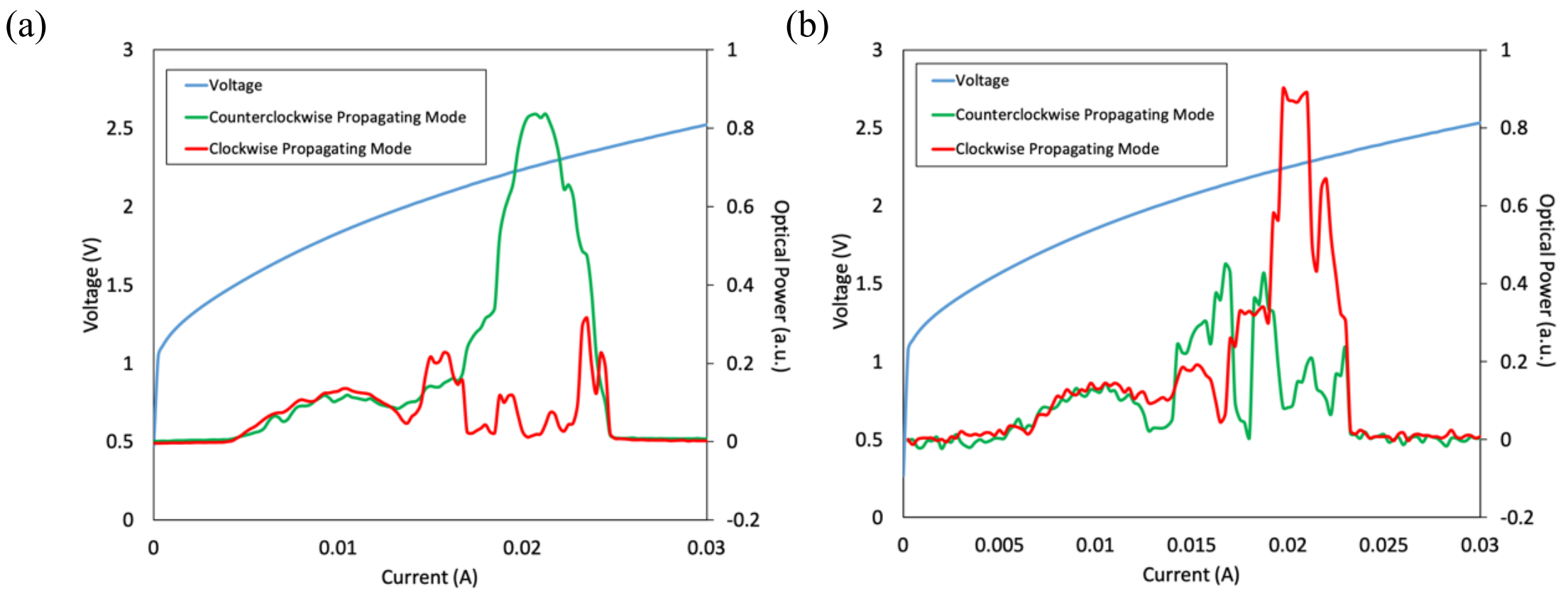

**Fig. 5 |** LIV curve of the memristor laser in the (a) LRS and (b) HRS.

The shift in the resonant wavelength is directly proportional to the thermo-optic coefficient of the oxide material, which is dependent on the temperature, as can be seen in Equation (1).

$$\Delta\lambda_{\mathrm{res}} = \frac{\lambda_{\mathrm{res}}}{n_{\mathrm{g0}}}\left(\frac{\partial n_{\mathrm{eff}}}{\partial T}\mathrm{d}T + n_{\mathrm{g0}}\frac{1}{L}\frac{\partial L}{\partial T}\mathrm{d}T\right) \tag{1}$$

Fabricated devices were tested on a copper stage with III-V side up with active temperature control. Current-voltage (I-V) curves were taken and shown in Fig. 4(a) and reveal a characteristic hysteresis curve consistent with a memristor operating in the bipolar mode. The current compliance was set to 10 microamps in the forward direction and 30 microamps in the reverse direction and the voltage was swept and sampled at a step of 0.2 V. Typically, the device had well below 1 nA of DC leakage current in the HRS state due to high quality of the $Al_2O_3$. The current was monitored as the voltage was swept from 0 to 20 V and the device was switched from the LRS to the HRS at around 16 V. The voltage was then swept in the opposite direction to -20 V to switch the memristor back to the HRS, which occurs at around -14 V. The total voltage distributes over the $Al_2O_3$ layer, semiconductor layers (n-GaAs, p-Si), and metal/semiconductor contact layers. Therefore, reducing the thickness of the $Al_2O_3$ layer, optimizing the semiconductor layers' doping concentrations and thicknesses, and improving the quality of the metal/semiconductor contact interface and the $Al_2O_3$ layer can help reduce the switching voltage[16]. The resistance of the memristor changes by about two orders of magnitude between the LRS and the HRS. Up to 4 states have been measured from a single laser with each resistance state has a corresponding output wavelength of the laser.

The endurance of memristive microring resonators made of the same materials have been previously measured and shown switching repeatability over $10^3$ cycles, implying that high endurance in these devices is similarly expected[17]. In addition, an Agilent B1500A semiconductor device analyzer with a pulse generator was used to set the memristor from the LRS to the HRS and measured the electrical switching speed to be less than 1 ns. The measured switching speed of these devices is three orders of magnitude faster than mainstream non-volatile flash memory and is comparable to memristor devices made of similar materials[18,19]. The electrical switching speed can be improved by applying a higher voltage pulse and by using a thinner oxide layer, a different oxide material, and by minimizing the series resistance semiconductor contact layers. Special-built waveguide structures can also be designed to extract high-speed electrical measurements without parasitic capacitive effects[20]. The optical switching speed of MOS microring lasers, on the other hand, is limited by the photon roundtrip lifetime inside the ring and the plasma dispersion effect. The bandwidth of similar III-V-on-Si MOS-type microring lasers have been demonstrated to be about 15 GHz, and III-V-on-Si MOS-type microring modulators to be about 30 GHz[21,22]. And due to the high thermo-optic coefficient of silicon, silicon microring modulators have also been integrated with heaters to demonstrate high wavelength tunability[23]. Since these memristor lasers can also be directly modulated by controlling the bias voltage of the memristor, data can be simultaneously stored, accessed, and sent with these devices at high speeds.

The measured optical output power came from two grating couplers at the ends of the same bus waveguide, i.e., output from clock-wise and counter clock-wise lasing directions. Due to structural symmetry, lasing can occur in the clockwise (CW) or counter-clockwise (CCW) direction in a traveling-wave laser cavity, e.g., ring laser here. CW or CCW lasing signal can be measured from coupling output from grating coupler either to the left or right hand side, respectively. Figs. 5(a) and (b) shows continuous-wave (cw) light-current-voltage (LIV) characteristics of a 50 µm-diameter laser at room temperature (RT) when the memristor was in the LRS and HRS at 5 V, respectively. After the memristor switches to the LRS, the lasing mode changed directions. This is likely due to wavelength-dependent optical reflections in the ring cavity or bus waveguide which alter the lasing direction when the memristor state and lasing wavelength are switched. Further work will be required to study how to control and take advantage of this lasing direction switching, which can be another useful property of this device in addition to wavelength switching.

Laser spectra were measured by sending output light from the grating coupler to an optical spectrum analyzer (OSA). The total loss coming from the grating coupler and waveguide loss was around 24 dB. All spectral testing was done by holding a constant laser injection current around 10-15 mA between P1 and N while changing the bias to memristor contact (P2-N) only. When the memristor laser is in the HRS, the lasing wavelength, $\lambda_{HRS}$, was blue-shifted by about 0.03 nm by sweeping the bias voltage across the memristor from 0 V to 5 V, as shown in Fig. 6(a). This process is dominated by the plasma dispersion effect, which agrees with observations made in heterogeneous InP-on-Si MOS microring devices[24]. After setting the memristor laser to the LRS, the lasing wavelength, $\lambda_{LRS}$, red-shifts by about 0.1 nm by sweeping the bias voltage across the memristor from 0 V to 5 V, as shown in Fig. 6(b), which indicates Joule heating through conductive paths inside the microring laser resonator[25]. The amount of wavelength shifting is directly proportional to the amount of current flowing through the memristor. We used a thermo-electric cooler (TEC) to heat up the device by 1.3°C when the memristor is in the HRS to achieve the same amount of laser wavelength red-shift of 0.1 nm and experimentally confirm the Joule heating effect caused by the memristor in the LRS. Thermal simulations confirm the shift in wavelength with the

(a) (b)

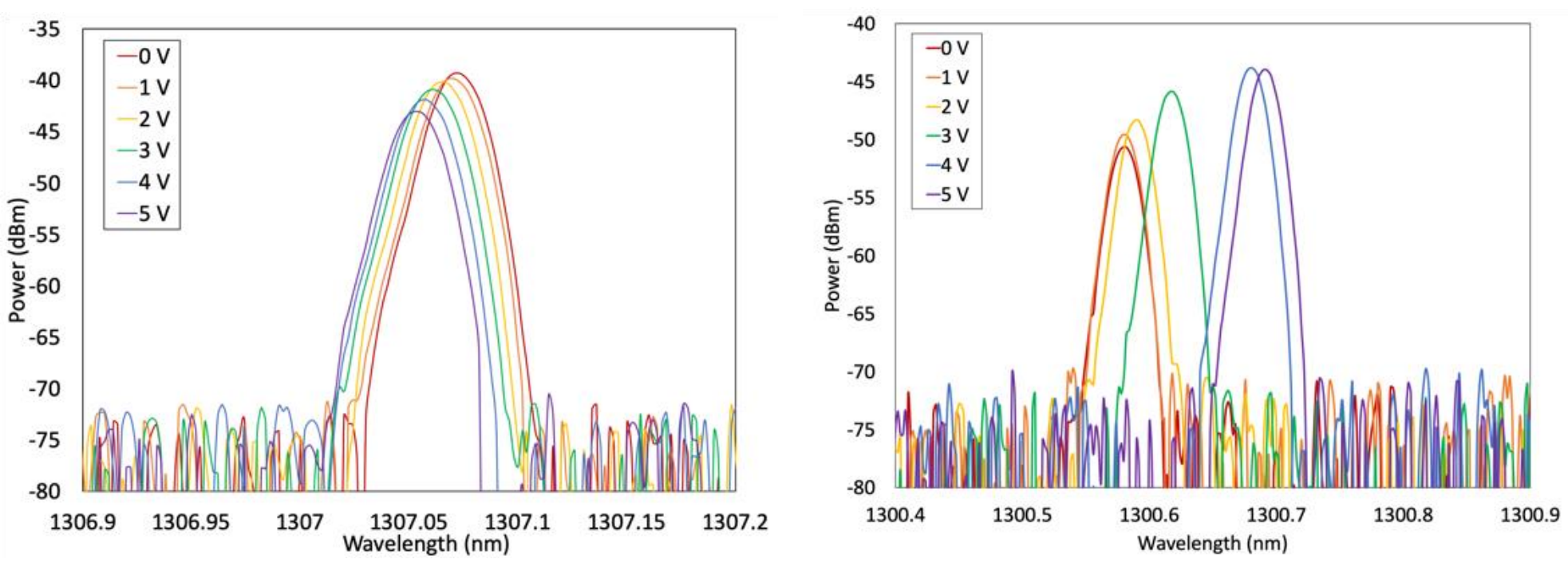

**Fig. 6 | a.** Measured laser output power in the HRS as a function of wavelength while biasing from 0 V to 5 V. Laser output wavelength blue shifts by 0.03 nm with increase in bias due to the plasma dispersion effect. **b.** Measured laser output power in the LRS as a function of wavelength while biasing from 0 V to 5 V. Laser output wavelength red shifts by ~0.1 nm with increase in bias due to localized Joule heating, and corresponding 1.3 °C increase in temperature of stage thermos-electric cooler (TEC).

proper corresponding Joule heating effect simulated in Fig. 2(b).

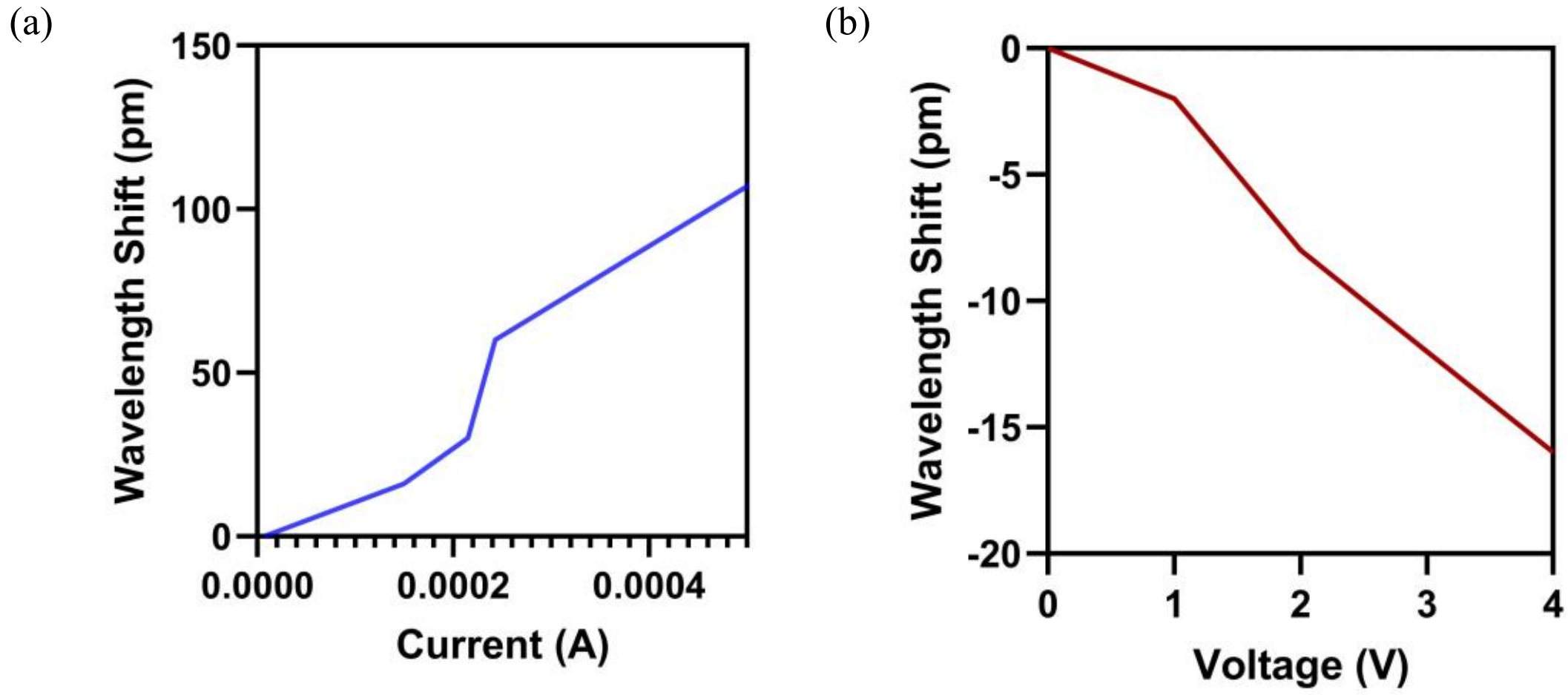

**Fig. 7 | a.** Wavelength shift as a function of current in memristor in the low resistance state. **b.** Wavelength shift

Future designs will feature device improvements and structural design changes such as oxide material and thickness in order to reduce set and reset voltages and increase switching speed. Reliability tests will be taken to measure endurance and yield and further investigated in order to design more robust devices. Also, TEM images will be taken to investigate the conductive filament formation within these devices and study the physical processes behind the switching mechanisms in these devices. These studies will aid in the design of future of devices such as in selection of oxide material. Furthermore, designs have been made and are currently being explored that integrate multiple memristor lasers together into circuits for multiple applications, such as in-memory computing.

In-memory processing offers a promising computing paradigm for future energy-efficient architectures which surpass the limitations of current von Neumann-based designs[26]. More specifically, the memristor laser can serve as a promising building block for photonic neuromorphic accelerators based on microring weight banks[27]. By controlling the output optical signals of memristor lasers with memristor storage and computing circuits, one can create a hybrid electronic-photonic artificial neural network, leveraging both the high-bandwidth, low-power capability of photonics and the high-density and re-programmability of electronics[28]. These devices can then drastically improve the performance and energy-efficiency of neuromorphic photonic computing circuits.

A memristive laser offers a key to optical data storage and optical memory chips, which can have numerous advantages to electronic counterparts[29]. For example, optoelectronic ternary content-addressable memory (TCAM) has gathered interest from the research community with the goal of replacing bulky and power-hungry electronic counterparts and reaching ultrafast memory read times[30]. These devices can also be programmed in real-time to interact with sensors in an external feedback system for optical field-programmable gate arrays (FPGAs) within an all-photonic digital computer[31]. Ultimately, since this device is readily compatible with a well-developed Si photonic technology platform, it opens up the possibility of integrating memory, computing, and high-speed optical interconnects all together on the same chip for the first time to advance in-memory computing and edge computing architectures[32].

**Acknowledgements**

The authors would like to thank Xiaoge Zeng and Stanley Cheung for their helpful discussions regarding the simulation of this device and Aidan Taylor for help in TEM imaging. The authors would also like to thank Thomas Van Vaerenbergh and Marco Fiorentino for their help with manuscript revision and high-level discussions.

**Author contributions**

B. T. conceived the idea, and led the simulations, experiments, and manuscript preparation. D. L. contributed to the device design and fabrication, manuscript preparation and project supervision. X. S. assisted electrical characterization of the memristors. J. S. and R. G. B. participated in the manuscript revision and high-level discussions.

**Competing financial interests**

The authors declare no competing financial interests.